\documentclass{PoS}

\title{Can the Supersymmetric Flavour Problem decouple in case of a Non Standard Supersymmetric Spectrum?}

\ShortTitle{Can the SFP decouple in case of a NSSS?}

\author{\speaker{Paolo Lodone}%
        \\
       Scuola Normale Superiore and INFN, Piazza dei Cavalieri 7, 56126 Pisa, Italy\\
       E-mail: \email{p.lodone@sns.it}}

\abstract{It has been shown that, in the context of the MSSM, the Supersymmetric Flavour Problem cannot be solved by just letting the sfermions of the first two generations be relatively heavy. The reason is twofold: naturalness of the Fermi scale on one side, need for positive squared stop masses on the other.
The situation is much more promising in models without a light Higgs boson, in which the goal can be met. The prices are: a relatively low messenger scale, semiperturbativity before the GUT scale, and some amount of degeneracy/alignement of order of the Cabibbo angle in the sfermion sector. A crucial role is played by the increased quartic coupling. The resulting phenomenology is quite different from the MSSM one in many respects\thanks{Review of \cite{Barbieri:2010pd}, to appear in the Proceedings of ICHEP 2010, Paris.}.}

\FullConference{35th International Conference of High Energy Physics - ICHEP2010,\\
		July 22-28, 2010\\
		Paris France}

\newcommand{\hhref}[1]{\href{http://arxiv.org/abs/#1}{\it arXiv:#1}}
\begin{document}

\section{Motivations}

One of the main sources of concern in the context of phenomenological supersymmetry is the lack of signals so far 
both in the Higgs and in the Flavour sectors.
Taking the view that the Supersymmetric Flavour Problem (SFP) may have something to do with a hierarchical structure \cite{HierSferm} of  sfermion masses, we argue about the possibility that ``the Higgs problem" and the SFP may be addressed at the same time by properly extending the MSSM \cite{Barbieri:2010pd}.

The motivation is once again naturalness (which is far from being a theorem), ie if we tolerate an amount of finetuning $1/\Delta$ \cite{finetuning} then the constraints:
\begin{eqnarray}
{m_{\tilde{t}}^2}/{m_h^2} \, \times \,
{\partial m_h^2}/{\partial m_{\tilde{t}}^2} &<& \Delta
\label{natbounds1} \\
{m^2_{\tilde{f}_{1,2}}}/{m_h^2} \, \times \, 
{\partial m_h^2}/{\partial m^2_{\tilde{f}_{1,2}}} &<& \Delta
\label{natbounds2}
\end{eqnarray}
may be both significantly reduced by pushing up the theoretical value of $m_h$.
%\footnote{Note that replacing the physical Higgs mass $m_h$ with the $Z$ mass or with any of the soft mass parameters for the Higgs doublets does not change the naturalness constraints on $m_{\tilde{t}}$ or on $m_{\tilde{f}_{1,2}}$, at least as long as the other physical Higgs bosons are not too close in mass to the lightest one, $h$, as we consider in the following for good phenomenological reasons. On this, see e.g. \cite{Barbieri:2006bg}.}.
Thus with some amount of degeneracy/alignement (besides the hierarchy) in the squark sector, one may achieve consistency with flavour observables without being unnatural (we will stick to $\Delta\leq 10$).

Another important bound on this ``decoupling" of the SFP comes from the conservation of colour and electric charge \cite{ArkaniHamed:1997ab}\cite{Agashe:1998zz}: we will see that a larger $m_h$ together with $M_{susy}\ll M_{GUT}$ helps in relaxing this bound too. It is also worth noticing that a low $M_{susy}$ alone would not be sufficient without further assumptions or more tuning.

Anticipating, we will discuss to what extent a ``Non Standard Supersymmetric Spectrum" (NSSS) with $m_h\gtrsim 200$ GeV and $m_{\tilde{f}_{1,2}} \sim 20\mbox{ TeV } \gg m_{\tilde{f}_3}$  can be natural and motivated.

\section{Hierarchical s-fermion masses and flavour physics}

Refering to \cite{Giudice:2008uk} for a recent analysis and to \cite{Barbieri:2010pd} for details, %\footnote{Notice that in that paper  one always considers $\delta_{LL} >> \delta_{RR}$ or viceversa.}
the situation can be summarized as follows.
Focussing on the squark sector, without degeneracy nor alignment the bounds on the masses of the first two generations are in the hundreds of TeV.
If we assume degeneracy and alignment of order of the Cabibbo angle, i.e. $\delta^{LL}_{12} \approx \frac{|m^2_1 - m^2_2|}{(m^2_1 + m^2_2)/2} \approx \lambda \approx 0.22$, and $\delta^{LL} \approx  \delta^{RR} >> \delta^{LR}$,
then the bounds become $O(10-100 \mbox{ TeV})$.
%to:
%\begin{equation}
%Real~\Delta S = 2 \Rightarrow  m_{\tilde{q}_{1,2}} \gtrsim 18~TeV   \, ,
%\end{equation}
%\begin{equation}
%Im~\Delta S = 2,~\sin{\phi_{CP}}\approx 0.3 \Rightarrow  m_{\tilde{q}_{1,2}} \gtrsim 120~TeV  \, .
%\end{equation}
Furthermore if
$\delta^{LL} >> \delta^{RR} , \delta^{LR}$
(or $\delta^{RR} >> \delta^{LL} , \delta^{LR}$), these bounds are replaced in the strongest cases by:
\begin{eqnarray}
\Delta C = 2 &\Rightarrow&  m_{\tilde{q}_{1,2}} \gtrsim 3~TeV  
\label{3TeV} \\
Im~\Delta S = 2,~\sin{\phi_{CP}} \approx 0.3 &\Rightarrow&  m_{\tilde{q}_{1,2}} \gtrsim 12~TeV 
\label{12TeV} 
\end{eqnarray}
 from CP conserving or CP violating effects respectively.
The exchange of the third generation of s-fermions may also produce too big flavour effects unless the off-diagonal $\delta_{i3}, i=1,2$ are small enough. Assuming $\delta^{LL}_{i3} \approx {m^2_{\tilde{f}_3}}/{m^2_{\tilde{f}_i}}$,
then a dominant constraint comes from $B-\overline{B}$ mixing:
\begin{equation}
\Delta B = 2 \Rightarrow  m_{\tilde{q}_{1,2}} \gtrsim  6~TeV ({m_{\tilde{q}_{3}}}/{500~GeV})^{1/2}.
\end{equation}
Similar or weaker constraints are obtained from the Electric Dipole Moments.

We conclude that, under relatively mild assumptions, $m_{\tilde{f}_{1,2}}\gtrsim 20$ TeV and $m_{\tilde{f}_3} \gtrsim 500$ GeV may be a way to solve the SFP.

%\begin{figure}
%\begin{center}
%\begin{tabular}{cc}
%\includegraphics[width=0.44\textwidth]{spettro2.pdf} &
%\includegraphics[width=0.44\textwidth]{1e5e6.pdf}
%\end{tabular}
%\caption{{\small \it Left: a representative NSSS with $m_h = 200\div 300$ GeV and $m_{\tilde{f}_{1,2}}\gtrsim 20$ TeV. Right: upper bounds on $m_h$ as function of the scale of semiperturbativity $\Lambda$ ($g_x^2, g_I^2, \lambda^2 = 4\pi$), for $U(1)$ (dotdashed, $\tan{\beta} >> 1$), $SU(2)$ (dashed, $\tan{\beta} >> 1$), and $\lambda$SUSY (solid,  $\tan{\beta} =1$). In the $SU(2)$ case values of $m_h \gtrsim 270$ GeV are hardly compatible with naturalness and  the EWPT.}}
%\label{spettro2eMaxMhThreeModels}
%\end{center}
%\end{figure}

\section{Naturalness with and without a light Higgs boson}

\begin{figure}
\begin{center}
\begin{tabular}{cc}
\includegraphics[width=0.39\textwidth]{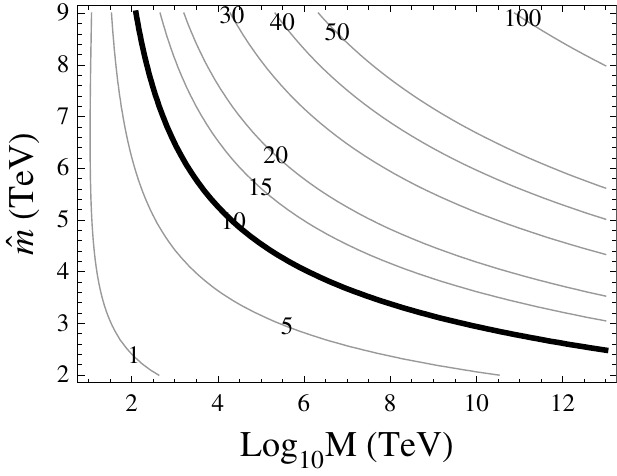} &
\includegraphics[width=0.41\textwidth]{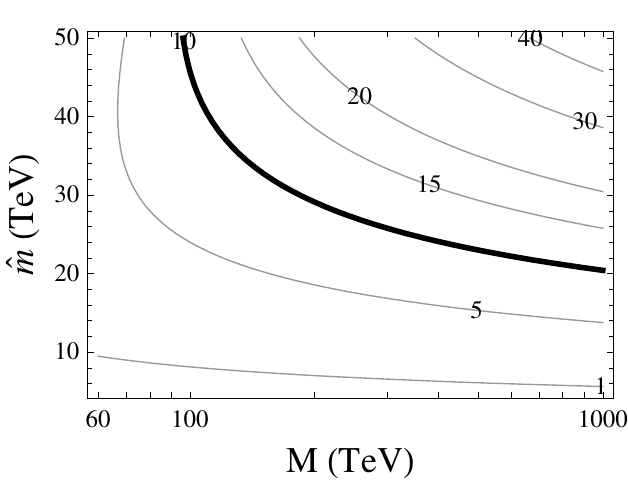}
\end{tabular}
\caption{{\small \it Upper bounds for different $\Delta$ on the masses of the $1^{st}$ and $2^{nd}$ generation scalars as function of the scale $M=M_{SUSY}$, assuming vertical degeneracy at $M$. Left: MSSM. Right: $\lambda$SUSY with $m_h = 250$ GeV.}}
\label{naturalnessMSSM}
\end{center}
\end{figure}

Fig. \ref{naturalnessMSSM} (left) shows the constraint \ref{natbounds2} in the MSSM as a function of the scale $M=M_{susy}$ at which the soft terms are generated, assuming a degenerate initial condition so that the Fayet-Iliopoulos term $Tr(Y\tilde{m}^2)$ vanishes and the running starts at two loop level. We immediatly see that even with very low $M$ we cannot satisfy the flavour bounds without a large amount of tuning (or without stronger assumptions about the flavour structure).

The situation is different in the context of extensions of the minimal model without a light Higgs boson \cite{nolightHiggsboson}\cite{Barbieri:2006bg}. Although we stick to a bottom-up point of view and allow the theory to become semiperturbative at a scale $\Lambda < M_{GUT}$, in order to mantain manifest consistency with the EWPT we require $\Lambda$ to be at least $5-10$ TeV, and this leads us not to consider raising the Higgs boson mass by means of higher dimensional operators \cite{effectiveoperators}.
Three very simple possibilities are then adding an extra $U(1)_x$ gauge group, enlarging the standard ElectroWeak gauge group to $SU(2)_I\times SU(2)_{II}\times U(1)_Y$, and $\lambda$SUSY which is the NMSSM with large coupling \cite{Barbieri:2006bg}.
Refering to \cite{Lodone:2010kt} for details and for a comparative study, we just notice that in all the three models it is possible to reach $m_h=200$ GeV at tree level.
%Fig. \ref{spettro2eMaxMhThreeModels} (right) shows the maximal value of $m_h$ in the three cases as function of the scale at which some coupling becomes semiperturbative. %, i.e. $g_x^2 =4\pi$ or $g_I^2 =4\pi$ or $\lambda^2 = 4\pi$ (the bounds in the gauge cases include 10$\%$ fine-tuning at most on the higher vev).
However, turning back to \ref{natbounds2}, it can be seen \cite{Barbieri:2010pd} that in the case of gauge extensions the naturalness bounds on $m_{\tilde{f}_{1,2}}$ are even stronger than in the MSSM.
On the contrary, in $\lambda$SUSY the first two generations of sfermions are not affected by the large $\lambda$ coupling, and a larger $m_h$ can indeed lower the level of finetuning. The result is shown in Fig. \ref{naturalnessMSSM} (right): we see that, for low enough values of $M$, the masses of the first two generations of sfermions can  go up to $20\div 30$ TeV in a natural way, thus going in the direction of solving the SFP via a NSSS.

\section{Constraint from colour conservation}

\begin{figure}
\begin{center}
\begin{tabular}{ccc}
\includegraphics[width=0.29\textwidth]{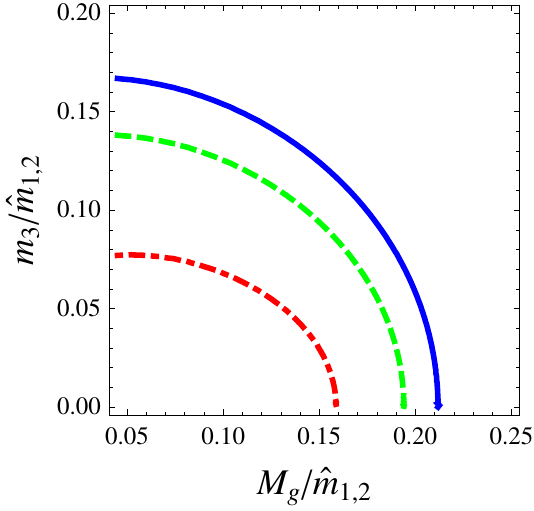} &
\includegraphics[width=0.28\textwidth]{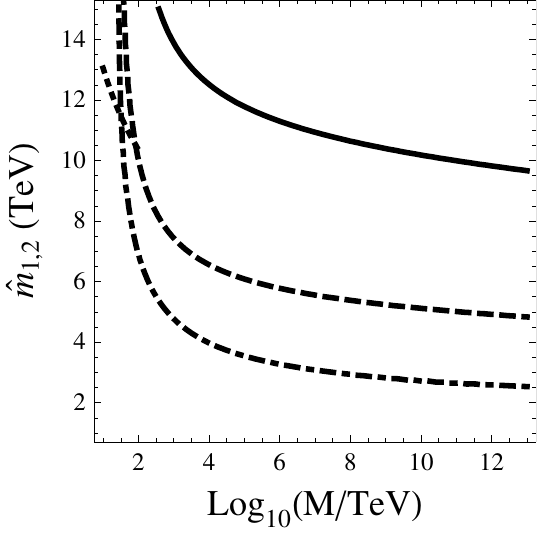} &
\includegraphics[width=0.29\textwidth]{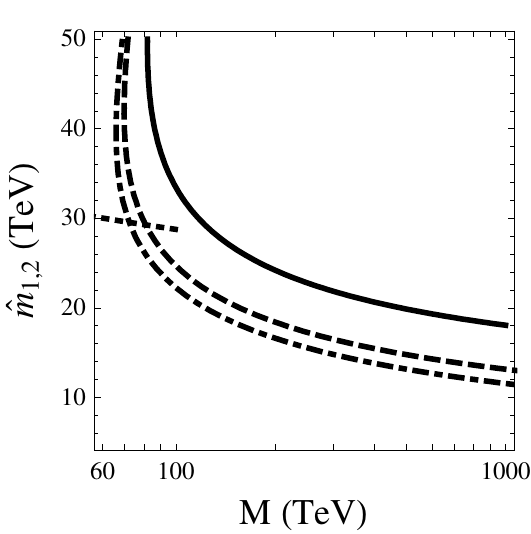} 
\end{tabular}
\caption{{\small \it Left: the regions below the curves are excluded, with ${M}/$TeV$=10^{13}$ (solid), $10^8$ (dashed), $10^3$ (dotdashed). Center: bound on $\hat{m}_{1,2}$ with ${M}_g/$TeV$=2$ (solid), 1 (dashed), $0.5$ (dotdashed) and $m_h=m_Z$. The dotted line below $M=100$ TeV estimates threshold effects (see text). Right: same for $m_h=250$ GeV.}}
\label{fig:colorunbreaking}
\end{center}
\end{figure}

As pointed out in \cite{ArkaniHamed:1997ab}, one has to check that the soft masses of the sfermions of the first two generations do not draw negative the squared masses of the lighter sfermions of the third one.
For example, neglecting the Yukawa couplings and focussing on $\tilde{Q}_3$ which gives the strongest bound of the squark sector, one has (up to two loops) with a degenerate initial condition $\hat{m}_{1,2}$:
\begin{eqnarray}
%\frac{d  m_{\tilde{u}_3}^2}{d \log \mu} &=& - \frac{1}{16\pi^2} \, \frac{32}{3} g_3^2 M_g^2 +\frac{8}{(16\pi^2)^2}    \left( \frac{16}{15} g_1^4 + \frac{16}{3} g_3^4 \right)\, \hat{m}_{1,2}^2  \label{u3evolution}
%\\
\frac{d  m_{\tilde{Q}_3}^2}{d \log \mu} &=& - \frac{1}{16\pi^2} \, \frac{32}{3} g_3^2 M_g^2 + \frac{8}{(16\pi^2)^2}   \left( \frac{1}{15} g_1^4 + 3 g_2^4 + \frac{16}{3} g_3^4 \right)\, \hat{m}_{1,2}^2   \label{Q3evolution} 
\end{eqnarray}
where we also neglected all the gauginos except the gluino.
The resulting bounds are quite strong if $M=M_{GUT}$, unless one chooses peculiar boundary conditions at $M$ so that the stability is protected \cite{peculiarboundarycond}.
One may instead take the view that they could be weakened by a lower $M$ scale. In  Fig. \ref{fig:colorunbreaking} (left) we report the analogous of Fig. 2 of \cite{ArkaniHamed:1997ab} for different values of $M$\footnote{The solid line is in agreement with \cite{ArkaniHamed:1997ab}, taking into account the fact that we keep only the gluino (while they keep all the gauginos with equal mass at $M_{GUT}$) and that our $M_g$ is the gluino mass at low energies.}.
Notice however that, if we insist on 10\% finetuning at most on the Fermi scale, the mere lowering of $M$ is not sufficient. In fact taking the value of $m_{\tilde{Q}_{3}}=m_3$ at $M$ which gives at most $10\%$ finetuning on the Fermi scale, one has\footnote{Which is valid both for the MSSM with large $\tan \beta$ ($m_h=m_Z$) and for $\lambda$SUSY with $\tan\beta\approx 1$ ($m_h \approx \lambda v$).} from \ref{natbounds1}:
\begin{equation}
%\frac{\partial \log v^2}{\partial \log m_3^2} \approx 
\frac{6 \, (m_t / \mbox{175 GeV})^2}{16 \pi^2} \, \frac{m_3^2}{m_h^2/2}  \, \log \frac{M}{\mbox{200 GeV}} \leq 10 \, ,
\end{equation}
and imposing that the running due to (\ref{Q3evolution}) does not drive $m_{\tilde{Q}_3}^2$ negative at $G_F^{-\frac{1}{2}}$ one obtains the bound shown in Figure \ref{fig:colorunbreaking} in the case of the MSSM (center) and in the case of $\lambda$SUSY with $\lambda v$= 250 GeV (right), as a function of $M$, $\hat{m}_{1,2}$, and the gluino mass at low energy ${M}_g$.
In the case $\hat{m}_{1,2}\sim M$ an important contribution comes from threshold effects, which can be estimated \cite{Agashe:1998zz} to give a bound $\hat{m}_{1,2}/m_{\tilde{Q}_3} \lesssim 25$. This estimate is shown as a dotted line in Figure \ref{fig:colorunbreaking}.

Thus the final conclusion is that also this constraint is satisfied in the case of interest, but notice that the increased quartic coupling of the Higgs plays a crucial role, allowing larger stop masses at $M$ with the same $10\%$ finetuning.

\section{Conclusions}

We gave attention to the possibility \cite{Barbieri:2010pd} that the Higgs mass problem and the SFP point towards extensions of the MSSM with a lightest Higgs boson naturally heavier than $m_Z$.
We discussed to what extent the SFP can be ``decoupled" by means of hierarchical soft mass terms without conflict with naturalness and colour unbreaking.
Among the phenomenological consequences of this NSSS, which may be useful to study more carefully, there are: abundance of top in the gluino decays which may give rise to distinctive signatures \cite{gluinotopphysics}, non standard features of the Higgs sector \cite{Barbieri:2006bg}\cite{Cavicchia:2007dp}, and a distinctive distortion of the relic abundance of the lightest neutralino \cite{Barbieri:2010pd} so that the LSP needs no longer to be ``well tempered" (\cite{ArkaniHamed:2006mb}).

%\section*{{Acknowledgments}}

%I thank Enrico Bertuzzo, Marco Farina, and especially Riccardo Barbieri.

%\appendix

%\section{Appendix}

\end{document}